# Participation-Sensitive Convergence and the Fragment First, Converge Later Pattern in Asynchronous Online Learning: A Topological Analysis Across 22 OULAD Courses

**Hitoshi INOUE[a*] & Koichi YASUTAKE[b]**
[a]*Nakamura Gakuen University, Japan*
[b]*Hiroshima University, Japan*
*jinoue@nakamura-u.ac.jp

**Abstract:** Asynchronous online learning offers temporal flexibility at a structural cost: learning communities tend to fragment rather than cohere. $\beta_0$, the number of connected components of a proximity graph over learner-week activity profiles, serves as a cohort-level indicator of this structure. Two questions remained unverified at scale: (1) does apparent $\beta_0$ convergence reflect genuine behavioral alignment or learner dropout? and (2) do assessment deadlines produce reproducible fragmentation–convergence cycles? We address both across all 22 OULAD courses (N > 22,000; 857 week-pairs). Changes in $\beta_0$ strongly co-vary with active learner changes (pooled r = 0.387; median per-course r_delta = 0.459, 20/22 courses), identifying $\beta_0$ as a participation-sensitive indicator: $\beta_0$ and active learner counts co-respond to deadline events rather than one causing the other. Deadlines produced fragmentation in 82.6% of assessments and the full Fragment First, Converge Later (FFCL) cycle in 60.2%. 3-phase analysis confirmed structural fragmentation as the dominant long-term trajectory (90.9% of courses), moderated by curriculum structure. These findings establish $\beta_0$ as a participation-sensitive structural indicator with direct implications for AI-augmented learning analytics design.

## 1. Introduction

AI-augmented learning analytics systems increasingly generate narrative explanations for educators—but when a cohort-level indicator is ambiguous, such narratives can systematically amplify misinterpretation at scale. $\beta_0$, the number of connected components of a proximity graph over learner-week activity profiles, is one such indicator: a declining $\beta_0$ can appear to signal growing cohesion, yet may instead reflect learner dropout reducing behavioral diversity. Without empirical grounding at scale, AI-generated narratives built on $\beta_0$ risk misleading educators in exactly the courses where dropout is highest. Dropout clusters around assessment deadlines in asynchronous online courses (Kizilcec et al., 2013), where completion rates remain below 15% (Jordan, 2014)—making this misreading particularly consequential.

$\beta_0$ — the number of connected components of a proximity graph over learner-week activity profiles, with consecutive weeks joined in a zigzag diagram of inclusions (Carlsson & de Silva, 2010) — tracks how learner groups split and merge over time. In a single-course pilot (DDD_2013B, N = 1,303), an unexpected pattern emerged: rather than bringing learners together, as constraint-dependent convergence theory predicts (Engeström, 2001), deadlines sharply increased $\beta_0$ — fragmentation rather than convergence (e.g., Week 7: +34.0%) —

Published in the Proceedings of ICLEA 2026: 2nd International Conference on Learning Evidence and Analytics. Copyright © Asia-Pacific Society for Computers in Education. Posted with permission under the rights retained by the authors, which permit non-commercial reproduction and distribution provided the authors receive credit and the proceedings are cited as the source of first publication. Available at: https://library.apsce.net/index.php/ICLEA/article/view/6320

followed by post-deadline declines co-occurring with falling participation. We call this two-stage pattern Fragment First, Converge Later (FFCL). It raises two questions: (1) does the post-deadline $\beta_0$ decline reflect genuine convergence, or does it simply reflect attrition reducing behavioral diversity (the learner attrition hypothesis)? and (2) is FFCL reproducible across courses, or an artifact of a single design?

This paper tests both questions systematically across all 22 OULAD courses. We make four contributions: (1) Methodological: we introduce the Union Zigzag construction and a systematic ε selection procedure via sensitivity analysis. (2) Attrition evidence: we quantify how much of the $\beta_0$ variation across 857 week-pairs is explained by attrition versus genuine behavioral alignment. (3) FFCL evidence: we establish how often the FFCL pattern occurs and characterize its properties across 22 courses. (4) Design guidance: we identify curriculum structure as a key factor shaping $\beta_0$ dynamics. We translate these findings into concise AI-robust guidance for dashboard narratives.

## 2. Background and Research Questions

Individual-level learning analytics models — such as dropout prediction (Chen et al., 2024) and transition networks (Saqr et al., 2026) — cannot tell us whether a learner cohort is breaking into isolated subgroups or consolidating into cohesive clusters. This distinction matters when designing interventions for the whole cohort. $\beta_0$ fills this gap by tracking collective behavioral structure over time. Stahl et al. (2006) argue that CSCL research must attend to group-level processes distinct from individual learning — and $\beta_0$ is designed precisely to capture that distinction. Unlike transition networks, which model individual behavioral sequences, Zigzag PH operates on the topology of the entire cohort's behavioral space at each time step, capturing collective structure rather than individual trajectories. The Union construction further detects learners becoming isolated from the main cohort as new $\beta_0$ components — directly operationalizing fragmentation in a way that sequence-based models cannot. Unlike aggregate dispersion measures such as engagement entropy or Gini coefficients, $\beta_0$ counts the number of distinct behavioral subgroups, making structural fragmentation directly interpretable rather than inferred from overall dispersion. Assessment distribution moderates learning strategy trajectories (Cristea et al., 2025), motivating RQ3 on how it shapes $\beta_0$.

**RQ1:** Is $\beta_0$ variation primarily associated with participation change (attrition-related context) rather than genuine behavioral convergence (learner attrition hypothesis)?
**RQ2:** Is the FFCL pattern (deadline-driven fragmentation followed by convergence) reproducible across 22 OULAD courses?
**RQ3:** How does curriculum structure (assessment distribution) moderate $\beta_0$ dynamics?

## 3. Method

### *3.1 Dataset and Preprocessing*

We used the Open University Learning Analytics Dataset (OULAD; Kuzilek et al., 2017). It covers 22 course presentations (N > 22,000 learners; 7 modules, 2013–2014) and includes VLE interaction logs, assessment records, and registration data. We extracted weekly activity counts across 10 VLE activity types, then z-standardized all features globally across the full course period to preserve temporal comparability for Zigzag PH.

### *3.2 ε Selection via Sensitivity Analysis*

We ran a sensitivity analysis over six values of ε (0.50–1.75) across the 20 courses within the target band, measuring the Union-graph isolation rate. ε = 1.00 met the 10–20% target in 11 of 20 courses (55.0%) and ε = 0.75 in 13 of 20 (65.0%). The series reported here was nevertheless computed at ε = 1.5, the value configured throughout the analysis pipeline, which this criterion does not select: at ε = 1.5 the mean isolation rate is 4.1% and no course falls within the target band. We adopted a single global ε to preserve cross-course comparability, and we report the threshold actually used rather than the one the criterion favours; the limitations below give the robustness of the conclusions to this choice.

### *3.3 $\beta_0$ Computation, Attrition Testing, and Statistical Analysis*

For each course, we computed weekly learner similarity matrices using Euclidean distance over 10-dimensional standardized features, then applied the Union construction (ε = 1.5) to obtain the connected components of each union complex directly, without computing persistence pairs, giving $\beta_0$ for each of the 857 week-pairs. Importantly, $\beta_0$ was computed over the learners active in each week, so each point in the graph is one learner-week and the point count varies with participation. ΔN (week-to-week change in weekly active learners, derived from VLE logs) is therefore not external to $\beta_0$: the two are entangled by the construction, which is what makes the attrition hypothesis worth testing rather than assuming. To test the learner attrition hypothesis, we computed two correlations per course: r_level = corr($\beta_{0,t}$, $N_t$) and r_delta = corr($\Delta\beta_{0,t}$, $\Delta N_t$), where t indexes weeks, $\beta_{0,t}$ is $\beta_0$ at week t, and $N_t$ is the active learner count at week t. Because $\beta_0$ is computed over the weekly-active learners, r_delta measures their co-movement at deadline events and does not identify a causal direction. Courses with r_delta ≥ 0.3 were classified as participation-sensitive (small-to-medium effect size threshold; p-values are indicative given within-course autocorrelation).

Because the point count is a function of participation, the association between the two series could in principle be mechanical. Recomputing $\beta_0$ over a fixed cohort — the learners present in every week of a course, so that the number of points cannot vary — removes that mechanism. Over the 20 courses where such a cohort exists, the median within-course r_delta falls from 0.453 to 0.380; restricting to cohorts of at least 10 learners gives 0.439 against 0.384, and at least 20 gives 0.460 against 0.399. Holding the point count fixed therefore leaves most of the association in place, so the varying point count cannot account for it. This is weaker than showing that the confounding is a property of the cohort's behavior, and it is as much as the comparison supports.

### *3.4 FFCL Pattern Detection*

Deadlines were selected as the primary externally imposed constraints (Kizilcec et al., 2013), with $\beta_0$ spikes confirmed to concentrate around assessment weeks over other course events. For each assessment deadline week d, we defined fragmentation as $\beta_{0,d}$ exceeding the mean $\beta_0$ of the preceding 3 weeks, and convergence as mean post-deadline $\beta_0$ (weeks d+1 to d+3) falling below $\beta_{0,d}$. A complete FFCL cycle requires both conditions. The FFCL rate is the proportion of assessments per course where both conditions hold. The 3-week window captures immediate post-deadline reorganization; results are directionally robust to modest shifts in window width (fragmentation rate: approximately 68% at W = 2, 3, and 4 weeks; FFCL rate: 52.6%, 47.6%, and 46.8% respectively; 17 of 22 courses maintained the same FFCL > 50% classification across all three windows). For the 3-phase analysis, we divided each course into equal thirds and classified the $\beta_0$ trajectory from Phase 1 to Phase 3 as either fragmenting or converging. Three phases represent the minimum granularity to distinguish early, middle, and late course structure; 2-phase (halves) analysis yielded consistent directional results.

## 4. Results

### *4.1 $\beta_0$ Patterns Across 22 Courses*

$\beta_0$ ranged from 4.6 (EEE_2013J) to 29.8 (DDD_2013B; Table 1). FFF courses showed the highest mean $\beta_0$ (18.0–22.5, coefficient of variation CV = 1.40–1.46, peak 123–157); EEE courses showed consistently low $\beta_0$ (4.6–10.3, CV = 0.88–1.11, no extreme spikes), providing initial evidence that curriculum design shapes collective behavioral structure.

Table 1. *$\beta_0$ patterns across course types: mean, variability, and dominant mechanism*

| Course Type | n | Mean $\beta_0$ (range) | CV (range) | Peak $\beta_0$ | Dominant mechanism |
|---|---|---|---|---|---|
| EEE | 3 | 4.6 – 10.3 | 0.88 – 1.11 | ≤40 | Mixed / Convergence |
| DDD | 4 | 14.6 – 29.8 | 0.67 – 1.04 | 60–85 | Mixed |
| FFF | 4 | 18.0 – 22.5 | 1.40 – 1.46 | 123–157 | Participation-sensitive |
| AAA/BBB/CCC/GGG | 11 | 6.5 – 25.0 | 0.64 – 1.31 | Varied | Mixed |

## *4.2 Learner Attrition Hypothesis (RQ1)*

The median r_delta across 22 courses was 0.459, well above the 0.3 threshold. Using a stricter threshold for dominance (r_delta > 0.5), 6 courses (27%) were participation-sensitive dominant, 14 (64%) mixed, and only 2 (9%) showed convergence dominance (EEE_2013J, DDD_2014J). Supplementary pooled r = 0.387 (835 week-pairs, first week per course excluded where ΔN is undefined, $p < 0.0001$); the primary evidence — per-course r_delta — shows that $\beta_0$ changes and active learner count changes co-respond consistently across courses. All 4 FFF courses showed r_delta > 0.5 (range 0.51–0.62, all $p < 0.003$), while r_level was near zero (range −0.09 to +0.03). This dissociation shows that $\beta_0$ spikes co-occur with periods of falling participation, not merely large cohort size.

## *4.3 FFCL Pattern Detection (RQ2)*

Figure 1 contrasts two representative courses: the FFF-type course (FFF_2013B) shows repeated large fragmentation spikes driven by concentrated Computer-Marked Assessment (CMA) deadlines, while the EEE-type course (2013J_EEE) shows moderate, stable $\beta_0$ responding to evenly distributed Tutor-Marked Assessment (TMA) deadlines. Both exhibit the FFCL pattern at individual deadlines, but temporal structure and magnitude differ markedly.

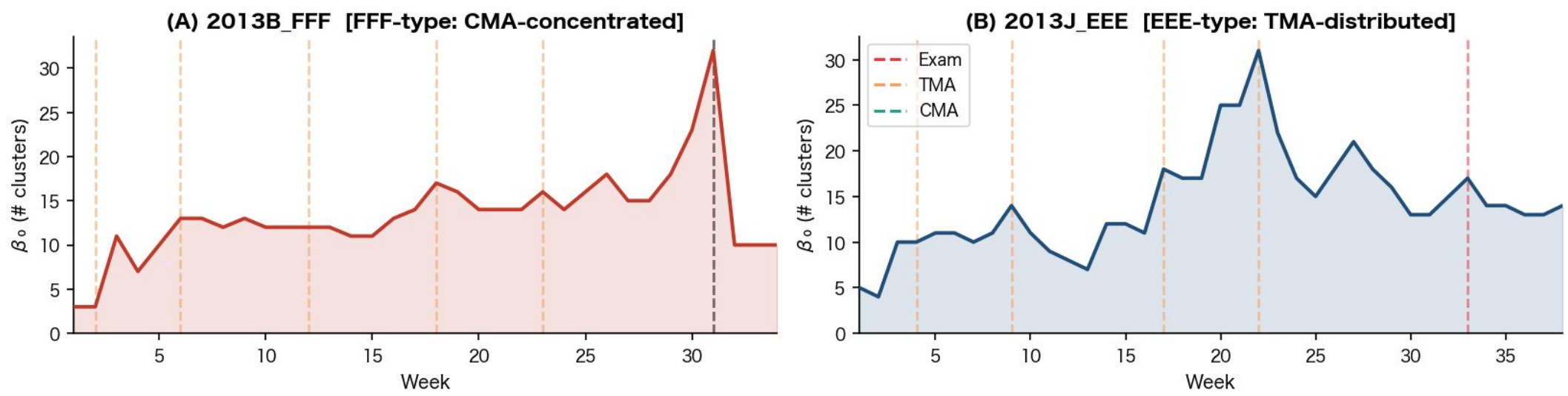

*Figure 1.* Representative $\beta_0$ trajectories contrasting curriculum structure types

Figure 2(A) and Table 2 present the FFCL results. Across 22 courses, deadline weeks produced $\beta_0$ increases (fragmentation) in an average of 82.6% of assessments (SD = 13.8%, range 54.5%–100.0%). The complete FFCL cycle appeared in an average of 60.2% of assessments per course (SD = 20.3%, range 28.6%–100.0%). Thirteen of 22 courses (59.1%) exceeded a 50% FFCL rate.

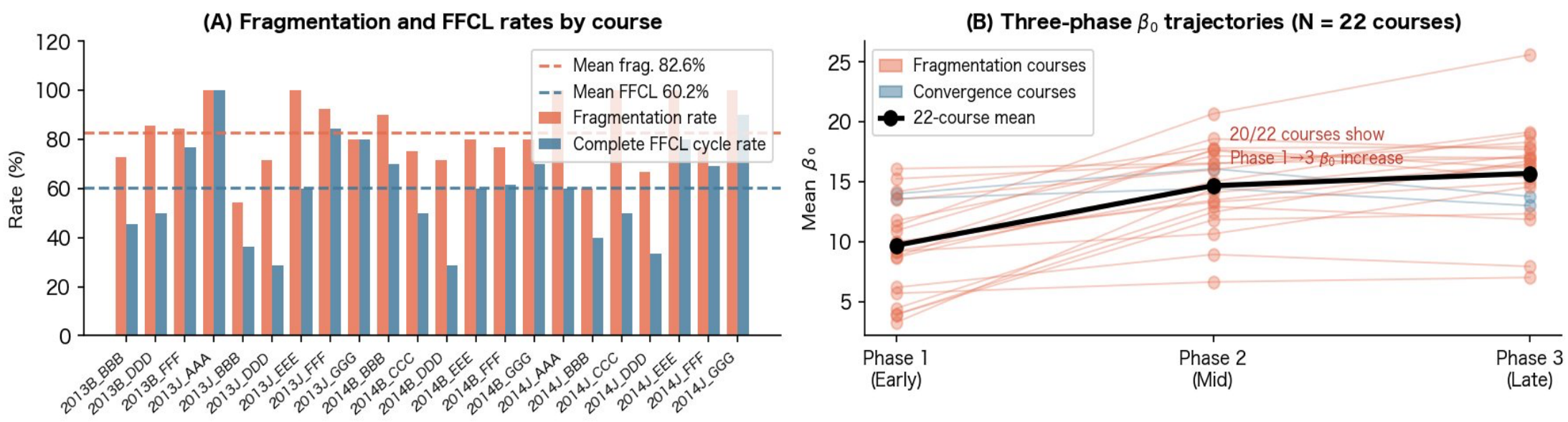

*Figure 2.* FFCL pattern validation across 22 OULAD courses

Table 2. *FFCL pattern detection and 3-phase $\beta_0$ analysis (selected courses and 22-course summary)*

| Course | N assess. | Frag. % (SD) | FFCL % (SD) | P1 $\beta_0$ | P3 $\beta_0$ | Δ% (P1→P3) |
|---|---|---|---|---|---|---|
| 2013B_FFF | 13 | 84.6 | 76.9 | 9.91 | 16.42 | +65.7% |
| 2013B_BBB† | 11 | 72.7 | 45.5 | 14.0 | 13.8 | −1.6% |
| 2013J_BBB† | 11 | 54.5 | 36.4 | 13.6 | 13.0 | −4.3% |
| 22-course mean | — | 82.6 (13.8) | 60.2 (20.3) | 9.8 | 15.6 | +92.4% |

*†Courses with overall convergence (P1→P3 $\beta_0$ decrease). Parentheses indicate SDs for the 22-course means. Across 22 courses, the median $\Delta\beta_0$ during deadline-week fragmentation is +2.0. The median post-deadline $\Delta\beta_0$ (mean of weeks d+1–d+3 vs. week d) is –1.0; 62.7% of post-deadline windows indicate convergence. P1 = Phase 1; P3 = Phase 3.*

3-phase analysis, shown in Figure 2(B), revealed that 20 of 22 courses (90.9%) had higher $\beta_0$ in Phase 3 than in Phase 1, with a mean increase of +92.4% (SD = 110.3%). Only 2 courses — BBB_2013B (−1.6%) and BBB_2013J (−4.3%) — showed marginal overall convergence. One-way ANOVA comparing $\beta_0$ across phases was significant in 16/22 courses ($p < .05$), confirming that these temporal trajectories are not random. Structural fragmentation, not convergence, is the dominant long-term pattern in asynchronous online learning.

### *4.4 Statistical Deadline Effects*

Mann-Whitney U tests confirmed positive $\beta_0$ change rates at deadline weeks in 18/22 courses (81.8%; mean = +13.8%). No course reached FDR-corrected significance (all p_FDR > 0.05); we treat these as supplementary evidence, with primary support from the directional consistency across courses (82.6% fragmentation rate). The directional consistency significantly exceeds chance: across all 130 deadline-week assessments pooled, 90 (69.2%) showed $\beta_0$ increases relative to the pre-deadline baseline, exceeding chance (binomial test: $p < 0.001$).

## 5. Discussion

### *5.1 Reframing $\beta_0$ as a Participation-Sensitive Structural Indicator*

The primary pattern — r_delta > 0.3 in 91% of courses, pooled r = 0.387 (supplementary; per-course median = 0.459) — establishes $\beta_0$ as a participation-sensitive indicator whose apparent convergence co-varies with active learner changes. $\beta_0$ is computed over the learners active in each week, so the point count varies with participation and the two series are entangled by construction; "participation-sensitive" names that entanglement together with the measured co-movement at deadline events, not a demonstrated causal direction (see Limitations). The fixed-cohort control reported in Section 3.3 is what shows the association is not merely mechanical: holding the point count fixed leaves most of it in place. The observed Phase 1→3 $\beta_0$ increase in 90.9% of courses is therefore a property of the weekly-active population rather than evidence about persistent learners specifically. In AI-augmented analytics, this ambiguity is amplified: a dashboard or AI assistant that frames a declining $\beta_0$ as "growing cohesion" can spread a systematically misleading narrative. $\beta_0$ should be interpreted alongside active learner count data. The 9% of courses showing convergence dominance (EEE_2013J, DDD_2014J) suggest conditions where genuine community formation occurs — supporting the hedged interpretation architecture of the TopoLA Dashboard (Inoue & Yasutake, 2026).

### *5.2 Implications for Intervention and Curriculum Design*

The 3-phase result — 90.9% of courses show increasing $\beta_0$ — has a direct implication: without active intervention, asynchronous learning communities fragment progressively over time. The post-deadline convergence window (the weeks immediately after a deadline) is the structural moment of lowest $\beta_0$ and therefore the optimal time for community-oriented scaffolding. For curriculum designers, assessment distribution — not just difficulty — is a key design variable. FFF's late CMA concentration creates predictable fragmentation crises that EEE's distributed schedule avoids; more even assessment spacing would substantially reduce $\beta_0$ spike amplitude and structural collapse. Learning analytics systems can monitor $\beta_0$ trajectories in real-time to flag these critical intervention windows. From an Activity Theory perspective (Engeström, 2001), our findings reveal that deadlines fragment cohorts before synchronizing them — identifying boundary conditions under which constraint-dependent coordination requires distributed assessment schedules.

From an AI-augmentation standpoint, AI-generated interpretations should be constrained by simple guardrails: (i) co-report $\beta_0$ with contemporaneous enrollment change, (ii) use hedged templates that flag participation change alongside $\beta_0$ decline rather than asserting cohesion, and (iii) treat deadline weeks as fragmentation-risk periods and post-deadline weeks as the optimal window for community scaffolding.

### *5.3 Limitations*

Six limitations deserve acknowledgment. First, OULAD represents a single institution; replication in MOOCs, K–12, or non-UK contexts is needed. Second, $\beta_0$ is computed over the learners active in each week, so low-activity learners enter and leave the graph and the point count varies with participation. The fixed-cohort recomputation reported in Section 3.3 shows that this variation does not account for the association, but the fixed-cohort series may itself carry an initial-homogeneity bias, since persistent learners may be more cohesive at the outset. Third, OULAD module codes are anonymized, so disciplinary differences may confound curriculum structure effects. Fourth, the series reported here was computed at $\varepsilon$ = 1.5, which the isolation-rate criterion of Section 3.2 does not select. Recomputing across $\varepsilon$ = 0.75–1.5 yields the same directional conclusions: the median per-course r_delta lies between 0.459 and 0.480, and 20 to 21 of the 22 courses exceed the 0.3 threshold at every value. Fifth, in courses with closely spaced deadlines (e.g., FFF-type), pre/post windows may overlap; window-width robustness (W = 2–4) partially mitigates this. Sixth, while the post-deadline window is identified as structurally optimal for intervention, the present study does not test the effectiveness of any specific scaffolding approach.

## 6. Conclusion

Across all 22 OULAD courses (N > 22,000; 857 week-pairs), this study establishes three findings: (1) $\beta_0$ is a participation-sensitive indicator — its changes and active learner count changes co-respond to deadline events (r_delta > 0.3 in 91% of courses; per-course median r_delta = 0.459); (2) FFCL is reproducible across courses (82.6% fragmentation rate; 60.2% full cycle rate); and (3) structural fragmentation is the dominant long-term trajectory (90.9% of courses show Phase 1→3 $\beta_0$ increases), moderated by curriculum structure. Together, these findings establish $\beta_0$ as a participation-sensitive structural indicator whose deadline-cycle fluctuations reliably follow the FFCL pattern.

## Acknowledgements

This work was supported by JSPS KAKENHI Grant Numbers JP25K00845, JP25K21951, JP23K22315, and JP23K17619.

## References

Carlsson, G., & de Silva, V. (2010). Zigzag persistence. Foundations of Computational Mathematics, 10(4), 367–405.

Chen, X., Zou, D., Xie, H., & Cheng, G. (2024). A systematic review for MOOC dropout prediction from the perspective of machine learning. Interactive Learning Environments, 32(5), 1642–1655.

Cristea, T. S., Heikkinen, S., Snijders, C., Saqr, M., Matzat, U., Conijn, R., & Kleingeld, A. (2025). Dynamics of self-regulated learning: The effectiveness of students' strategies across course periods. Computers & Education, 228, 105233.

Engeström, Y. (2001). Expansive learning at work: Toward an activity-theoretical reconceptualization. Journal of Education and Work, 14(1), 133–156.

Inoue, H., & Yasutake, K. (2026). TopoLA Dashboard: Making topological learning analytics accessible for exploring learner behavioral structure. Interactive Tools and Demos at ISLS 2026 (in press).

Jordan, K. (2014). Initial trends in enrolment and completion of massive open online courses. The International Review of Research in Open and Distributed Learning, 15(1), 133–160.

Kizilcec, R. F., Piech, C., & Schneider, E. (2013). Deconstructing disengagement: Analyzing learner subpopulations in massive open online courses. Proceedings of LAK 2013, 170–179.

Kuzilek, J., Hlosta, M., & Zdrahal, Z. (2017). Open University Learning Analytics dataset. Scientific Data, 4, 170171.

Saqr, M., Helske-Kunowski, S., Havu, M., & López-Pernas, S. (2026). Transition and frequency transition network analysis. In S. López-Pernas & M. Saqr (Eds.), Advanced Learning Analytics Methods. Springer.

Stahl, G., Koschmann, T., & Suthers, D. (2006). Computer-supported collaborative learning: An historical perspective. In R. K. Sawyer (Ed.), Cambridge handbook of the learning sciences (pp. 409–426). Cambridge University Press.